\documentclass[aps,twocolumn]{revtex4}
\usepackage{epsfig}
\newcommand{\be}{\begin{equation}}
\newcommand{\ee}{\end{equation}}

\begin{document}

\title{Critical Behaviour of the Drossel-Schwabl Forest Fire Model}

\author{Peter Grassberger}

\affiliation{John-von-Neumann Institute for Computing, Forschungszentrum J\"ulich,
D-52425 J\"ulich, Germany}

\date{\today}

\begin{abstract}
We present high statistics Monte Carlo results for the Drossel-Schwabl 
forest fire model in 2 dimensions. They extend to much larger lattices (up to 
$65536\times 65536$) than previous simulations and reach much closer to the 
critical point (up to $\theta \equiv p/f = 256000$). They are incompatible 
with all previous conjectures for the (extrapolated) critical behaviour, 
although they in general agree well with previous simulations wherever they 
can be directly compared. Instead, they suggest 
that scaling laws observed in previous simulations are spurious, and that 
the density $\rho$ of trees in the critical state was grossly underestimated.
While previous simulations gave $\rho\approx 0.408$, we conjecture that $\rho$ 
actually is equal to the critical threshold $p_c = 0.592\ldots$ for site 
percolation in $d=2$. This is however still far from the densities reachable 
with present day computers, and we estimate that we would need many orders 
of magnitude higher CPU times and storage capacities to reach the true critical
behaviour -- which might or might not be that of ordinary percolation.
\end{abstract}

\maketitle

\section{Introduction}

Empirical analyses suggest that power laws with anomalous exponents are ubiquitous
in nature, ranging from $1/f$ noise and earth quake distributions to fractal 
coast lines, species extinction rates, weather records, and the statistics of DNA 
\cite{mandelbrot,bak,buldyrev,koscielny}. On the other hand, it is well known 
that such scaling laws -- most clearly seen by linear relationships in log-log 
plots -- can be spurious. Log-log plots have a notorious tendency to suggest 
linear curves, even if there are no real power laws. 

It would thus be extremely useful if these empirical observations could be 
backed by theoretical models where power laws can be either proven exactly or 
at least verified beyond doubt by high statistics simulations. Unfortunately, 
equilibrium systems in general show anomalous scaling laws only at critical 
points which are codimension one phenomena: One has to fine tune some parameter 
(e.g. temperature) to reach them, otherwise no power laws are obtained. Thus, 
they cannot be used to justify why such power laws should be seen in nature.

A possible solution of this puzzle was indicated in \cite{btw,bak} where it was 
suggested that many {\it non-equilibrium} systems could be driven by their dynamics
into a critical state. The main ingredients of this {\it self organized criticality}
(SOC) is slow driving towards some instability and a mechanism to relax the 
tensions built up by the drive locally and partially. Since the tensions are not 
relaxed completely (in many models they are just redistributed), the state becomes 
marginally stable and apt to relaxation events on wider and wider length scales,
which then lead to anomalous scaling laws. The paradigmatic model is the 
``sand pile" of \cite{btw} which does not describe real sand piles but which 
was proven exactly to show anomalous scaling laws at least for some of its 
observables \cite{dhar,priezzhev,glkp}.

Another model which was proposed to show SOC is the forest fire model introduced 
independently by Henley \cite{henley} and by Drossel and Schwabl \cite{ds}. In this 
lattice model with discrete time each site can be either occupied by a tree or empty.
New trees are grown with small fixed rate $p$ on empty sites, and with a rate $f\ll
p$ sites are hit by lightning strokes which then burn the entire cluster of trees
connected to this site. Burning happens infinitely fast, so the only relevant 
parameter is the ratio $\theta = p/f$ which also sets the scale for the average 
fire size (the number of trees burnt after one lightning). Criticality is observed 
in the limit $\theta \to \infty$.

The Drossel-Schwabl model (called DS model in the following)
is different from other SOC models in two ways.
\begin{itemize}
\item It involves not only the separation between 
two time scales (the slow build-up of stress and the fast relaxation), but 
involves {\it three} time scales: The fast burning of connected clusters of trees,
the slow re-growth, and the even slower rate of lightnings.
\item The growth of trees does not lead to a state which is inherently 
unstable (as does the addition of sand grains to the top of a pile does), but only
to a state {\it susceptible} to being burnt. Without lightning, the tree density 
in any patch of forest can go far beyond criticality. When the lightning strikes 
finally, the surviving trees have a density far above critical.
\end{itemize}

Indeed, it was observed in \cite{grass} that the stationary steady state of 
the model is not ``critical" in most regions, in the sense that its tree density 
is not marginal for the spreading of fire. Rather, it is composed of large
patches of roughly uniform density, most of which are either far below or far 
above the critical density for spreading. Nevertheless, power laws were observed 
for several observables, partly because these patches occur with all sizes, 
so that also the fires had a broad spectrum of sizes.

While normal scaling with mean field exponents had been seen in \cite{ds}, all 
subsequent simulations \cite{grass,henley,christensen,clar,honecker} showed 
clear signs of anomalous scaling:
\begin{itemize}
\item The fire size distribution scaled, for small $s$ ($s$ is 
the number of trees burnt in one fire) and in the limit $\theta\to\infty$,
as $P(s) \sim s^{1-\tau}$ with $\tau \approx 2.15$;
\item For finite $\theta$, 
$P(s)$ is essentially cut off at $s_{\rm max} \sim \theta^\lambda$ with 
$\lambda \approx 1.1$; 
\item The average rms. radius $R(\theta)= \langle R^2\rangle^{1/2}$ of all 
fires scaled as $\theta^\nu$ with $\nu \approx 0.58$. Here, $R^2$ is the 
Euclidean distance of a burning tree from the site of lightning,
and the average is taken over all trees in all fires at fixed $\theta$; and
\item The rms. radius $R(s,\theta)$ of fires of fixed 
size $s$ scaled as $s^{1/D}$ where the fractal dimension $D$ of fires is
$D \approx 1.96$ \cite{clar} to $2.0$ \cite{grass}. 
\end{itemize}
Finally, the average 
density of trees was found to be $\rho = 0.408 - {\rm const}/\sqrt{\theta}$.

There were however seen already at that time large corrections to this scaling 
law and deviations from conventional ansatzes. Thus,
\begin{itemize}
\item The determination of $\tau$ was subject to large systematic
uncertainties \cite{grass};
\item The scaling $R \sim s^{1/D}$ was observed in \cite{grass} only for 
fires of typical size ($s\sim \theta$). For large $s$, $R$ was significantly 
larger; 
\item For finite $\theta$, the fire distribution $P(s)$ did not follow the 
normal scaling ansatz $P = s^{1-\tau} \phi(s/s_{\rm max})$ \cite{grass};
\item There are (at least) two differently divergent length scales 
\cite{honecker}: The correlation length evaluated from {\it all} pairs
of sites scales differently with $\theta$ than $R(\theta)$ \cite{footnote1};
\item Finite size behaviour is abnormal \cite{schenk0}.
\end{itemize}

In two recent publications, these problems were taken up again. In 
\cite{vespignani} it was claimed that they are due to non-leading 
corrections to scaling. In \cite{schenk} a more radical solution was 
proposed with two distinct classes of fires which both contribute to the 
scaling limit. In the latter paper also a connection to ordinary 
percolation was proposed, and the conclusion was backed by a ``coarse 
grained" model which supposedly could mimic the DS model at extremely 
large $\theta$ and on extremely large lattices.

It is the purpose of this paper to present very large simulations which 
show quite unambiguously that none of these describe really the true critical 
behaviour of the DS model. While our simulations agree perfectly with 
previous ones for the lattice sizes and $\theta$ values used there, we shall 
see that the supposed scaling laws are just transients. Even the present 
simulations do not reach the true asymptotic regime, but some suggestive 
features of the true asymptotics do emerge.

We describe our simulations and their results in the next section. In the 
last section we draw our conclusions.

\section{The Simulations}

Our simulations are straightforward, with a few subtleties. They follow 
Ref.\cite{grass} in taking $p\to 0$ by making no attempts to grow new trees
while a cluster burns. As in \cite{grass} we made {\it exactly} $\theta$
growth attempts on randomly chosen sites between two successive lightnings, 
instead of letting this number fluctuate. In the large $L$ limit this should 
not make any difference.  But in contrast to \cite{grass}, where a depth first
algorithm had been used, we used a breadth first algorithm to burn the cluster
\cite{footnote2}. 

In order to simulate very large lattices
with minimal storage, we use multi-spin coding, i.e. we use only one bit to 
store the status of a site. In this way we could simulate lattices of size 
$L\times L,\; L = 65536$ on computers with 1 GB main memory. Notice that we do 
not need to store for every tree whether it is burning or not, since the
burning trees are stored in a separate list which is updated at each time 
step. Boundary conditions were helical, i.e. sites are indexed by one scalar
index $i$, with neighbours having indices $i\pm 1$ and $i\pm L$, and with
$i+L^2\equiv i$. The largest previous simulations \cite{vespignani} had used 
$L=19000$.

We were careful to discard sufficiently long transients (between $1.6\time 10^6$ 
and $1.2\times 10^7$ lightnings; this is up to one order of magnitude longer 
than those in \cite{honecker,vespignani}) before taking statistics. This is 
needed since excessively large fires occur during these transients, and thus the 
tail of the distribution $P(s)$ is heavily influenced by them. We believe that 
previous analyses were affected by this problem which is easily overlooked 
since bulk properties (such as $\rho$ or $P(s)$ for typical $s$) show much 
faster convergence. The total number of fires in each run used for averaging was 
between $10^9$ (for $\theta\le 250$) and $9.3\times 10^6$ (for $\theta = 256000$; 
see Table 1). Previous authors \cite{vespignani} went only to $\theta = 32768$ 
with $2.5\times 10^7$ fires. Compared to that, our statistics is larger by 
roughly 1 order of magnitude.  All simulations were done on fast Alpha 
workstations. The CPU times per run varied between 15h and 4 weeks. As random 
number generator we used Ziff's four-tap generator with period $2^{9689}-1$ 
\cite{ziff}.

We tested for finite size effects by making for the same $\theta$ several 
runs on lattices of different sizes. Previously, finite size effects had been 
studied systematically in \cite{schenk0}, but only for much smaller lattices 
($L\le 2000$; the authors of \cite{vespignani} called their analysis a 
finite size analysis, but they actually made a conventional scaling analysis
without checking the finite size behaviour). For $\theta=64000$, e.g., we 
made runs with $L=2^{14}, 2^{15},$ and $2^{16}$. We verified that distributions 
like $P(s)$, $R(s,\theta)$, or $P(t)$ ($t$ is the burning time of a
fire) were independent of $L$ within the statistical accuracy, i.e. any 
systematic $L$-dependence was masked by statistical fluctuations. 

Systematic
dependencies were seen only for averaged quantities like $\langle s\rangle$,
$\rho$, or $R(\theta)$. Indeed, $\rho$ can be measured either immediately
before or immediately after a lightning. Since each lightning burns in 
average $\langle s\rangle$ trees (here $\langle s\rangle$ is averaged over 
{\it all} lightnings, whether they hit a tree or an empty site), we have 
$\rho_{\rm before} - \rho_{\rm after} = \langle s\rangle /L^2$. We found
that $\rho_{\rm before}$ decreased with $L$, while $\rho_{\rm after}$ 
increased with it (see Table 1). More precisely, we have 
\be
   \rho_{\rm after}(L) \approx \rho(\infty) - 0.6\,\langle s\rangle /L^2 .
\ee
Notice that $\langle s\rangle$ is given for $L\to\infty$ exactly by 
$\langle s\rangle = (1-\rho)\theta$ \cite{ds}. For finite $L$, one has 
to replace just $\rho$ by $(\rho_{\rm before}+\rho_{\rm after})/2$, to 
leading order in $1/L$. This was verified in the simulations, but it tests
just stationarity and the absence of gross mistakes.

The rms. radius of fires depended on $L$ and $\theta$ in a more complicated 
way, see Table 1. The data were less clear in that case, but they could be 
fitted by 
\be
   R(\theta)_{L=\infty} - R(\theta)_L \approx 0.004\,\theta^{3/2}/L.
\ee

\begin{figure}
\psfig{file=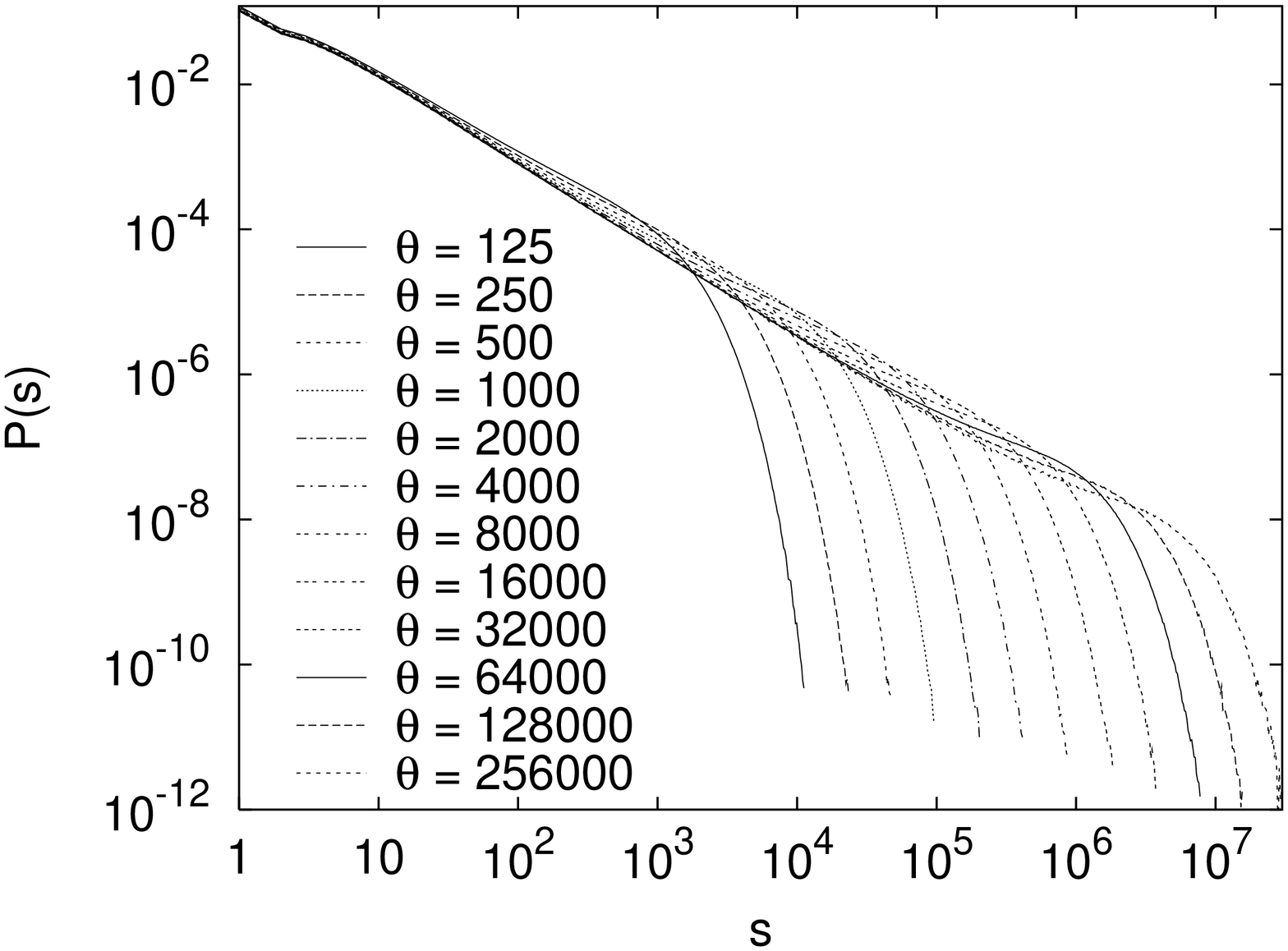,width=5.8cm,angle=270}
\psfig{file=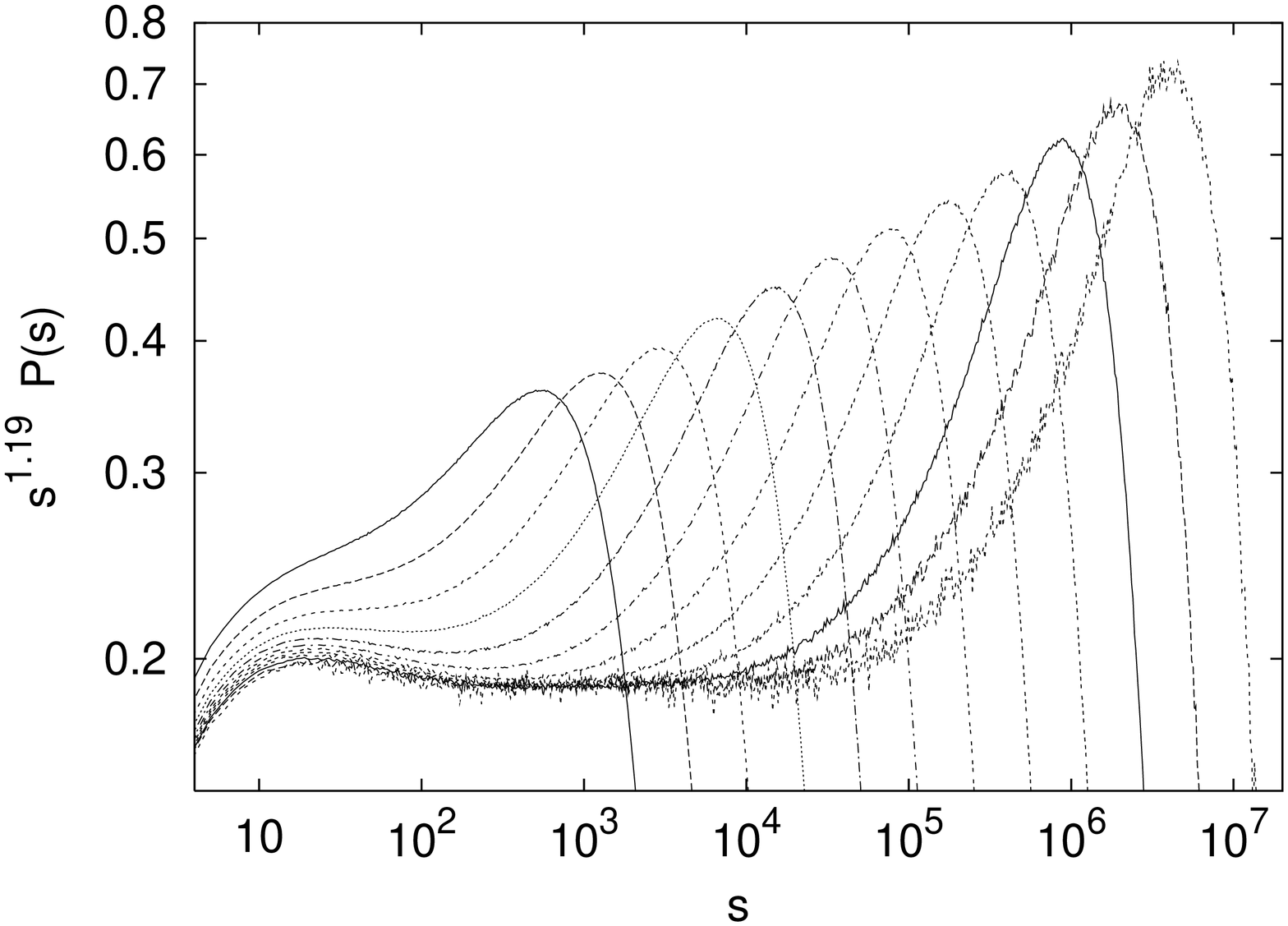,width=6.cm,angle=270}
\caption{(a) Log-log plot of $P(s)$ versus $s$, for fixed $\theta$ ranging 
   from 125 to 256000. Here and in Figs.2 and 5 we used logarithmic binning 
   with 2 per cent bin size, to suppress excessive statistical fluctuations.\\
   (b) Enlarged part of the same data, multiplied with $s^{1.19}$.}
\label{ps.ps}
\end{figure}

Results for the fire size distributions $P(s)$ are shown in Fig.1. We see the 
approximate power decay $s^{1-\tau}$ for $s < s_{\rm max}$ with $s_{\rm max}$ 
roughly proportional to $\theta$. But we see also the strong deviations from 
this power law first observed in \cite{grass}. Due to these deviations, $P(s)$ 
decreases with $\theta$ in the scaling region (making the effective value of 
$\tau$ increase with $\theta$), but has a growing bump at $s\approx s_{\rm 
max}$. In \cite{schenk} it was conjectured that the asymptotic value of $\tau$, 
estimated from the scaling region $1 \ll s \ll \theta$ in the limit $\theta 
\to \infty$, is $\tau=2.45$. This was based on heuristics and on simulations 
for small $\theta$ on very small lattices ($L=1300$).
In order to test it, we plotted $P(s)$ in Fig.1b after multiplying it with a 
suitable power of $s$. It is seen that $s^{1.19}\,P(s)$ becomes flat in a wider 
and wider region of $s$ as $\theta$ increases. For very small fires ($s < 300$)
$P(s)$ decreases faster than $s^{-1.19}$, but there is no indication that the 
slope increases with $\theta$ for $\theta > 8000$. Based on this evidence we 
would thus conclude that $\tau = 2.19\pm 0.01$, ruling thereby out the value of 
\cite{schenk} -- provided, of course, that there is no change of behaviour as 
$\theta$ becomes even larger.

\begin{figure}
\psfig{file=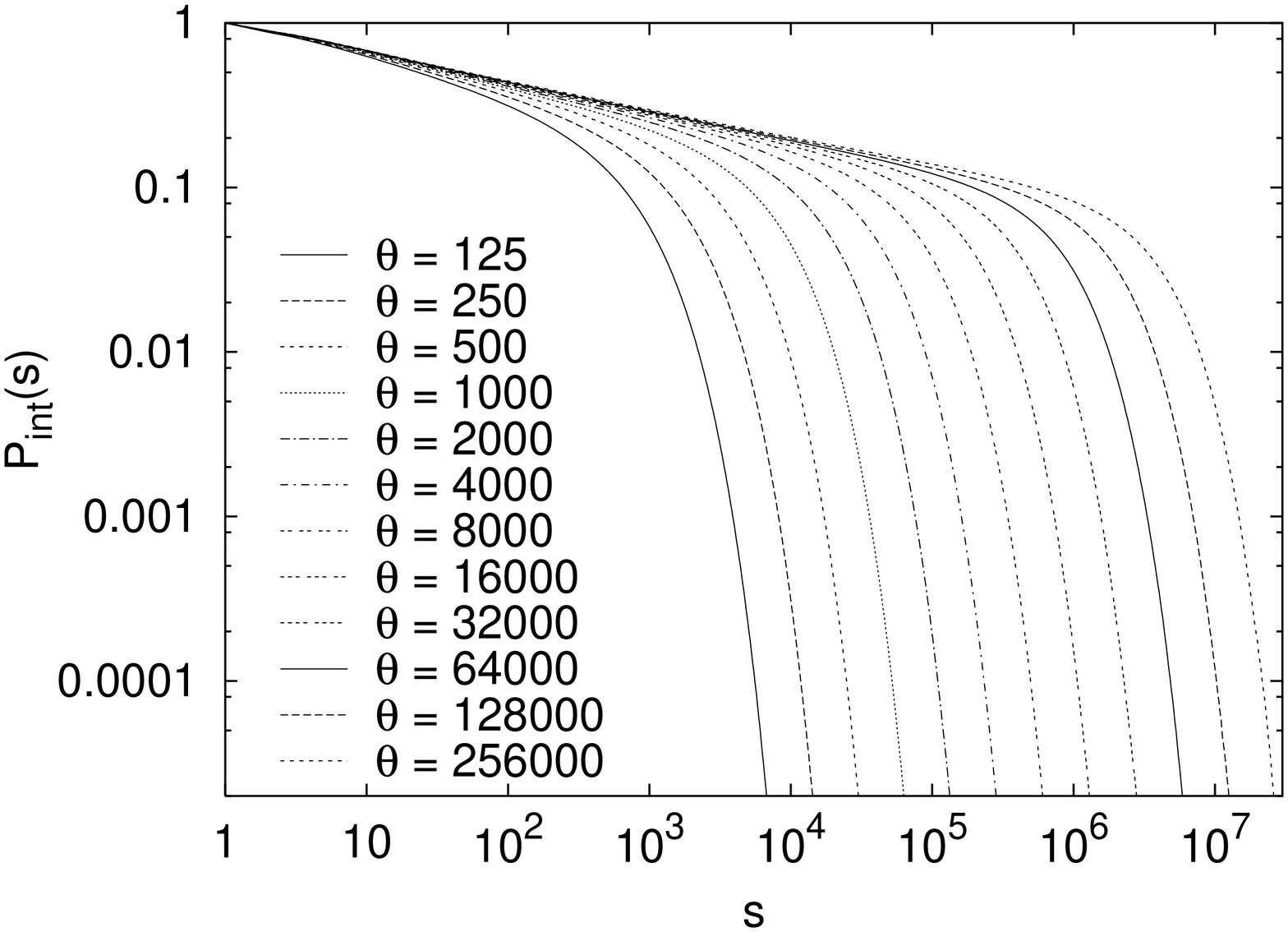,width=5.8cm,angle=270}
\psfig{file=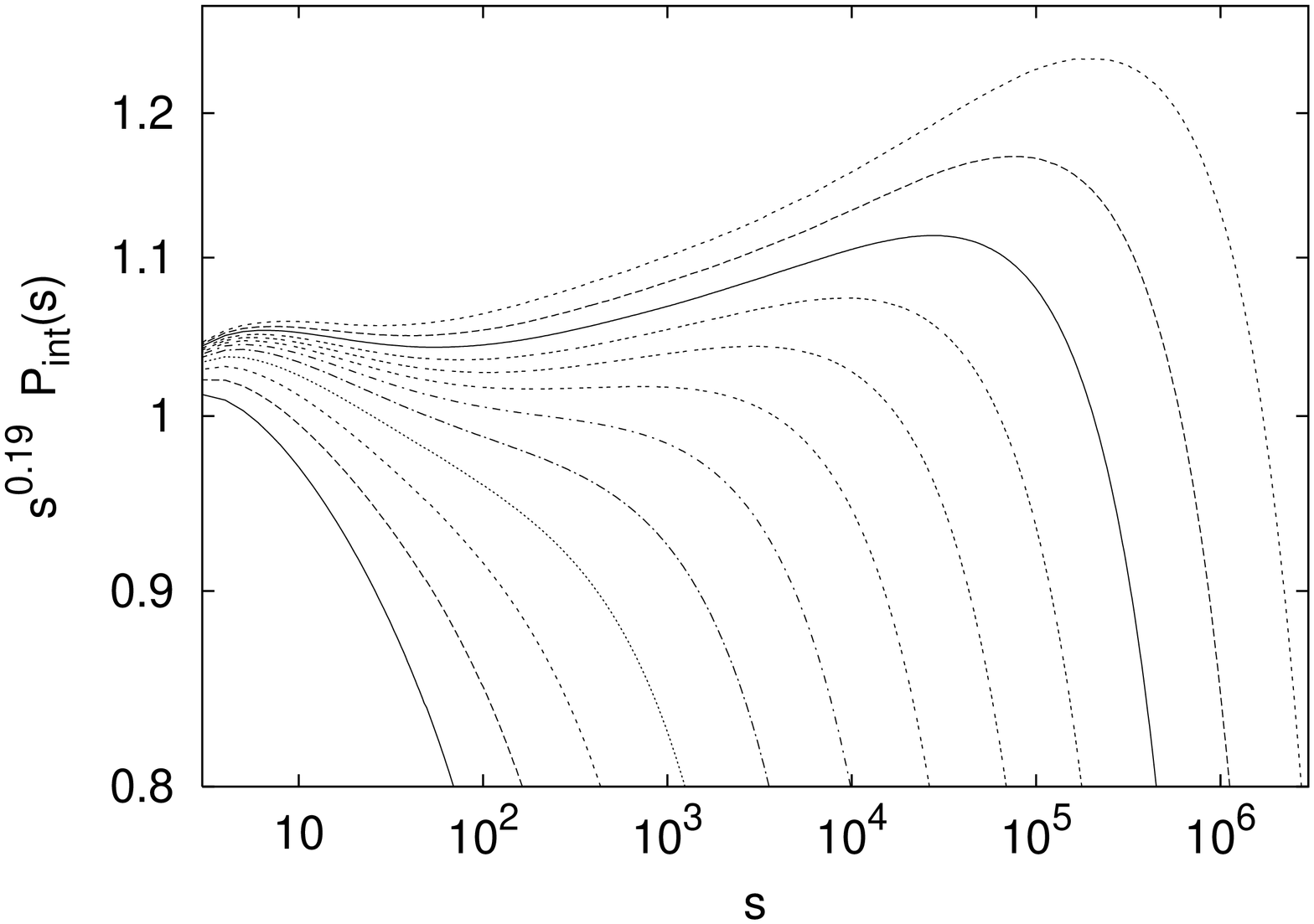,width=6.cm,angle=270}
\caption{(a) Log-log plot of $P_{\rm int}(s)$ versus $s$, for fixed $\theta$ 
   ranging from 125 to 256000. \\
   (b) Enlarged part of the same data, multiplied with $s^{0.19}$.}
\label{ps_int.ps}
\end{figure}

In \cite{grass} it was conjectured that the bump near $s\approx s_{\rm max}$ is 
due to the cut-off. If the {\it integrated} distribution $P_{\rm int}(s) = 
\sum_{s'=s}^\infty P(s')$ were just a power multiplied by a sharp cut-off, its 
derivative $P(s)$ would have a bump where the cut-off sets in. This bump would 
consist of those events which would have been in the tail which is cut off. If 
this were right and $P_{\rm int}(s)$ would indeed show normal scaling, we should 
expect the height of the bump in Fig.1b to be independent of $\theta$. But this 
is obviously not the case. Instead, it increases with $\theta$, suggesting a 
different scaling law $s^{1-{\tau'}}$ for the envelope of the curves in Fig.1a, 
with $\tau' = 2.111\pm 0.006$.

A similar conclusion is reached by looking directly at $P_{\rm int}(s)$.
Indeed, log-log plots of $P_{\rm int}(s)$ versus $s$ are much more straight 
in the scaling region (Fig.2a). But again a blow-up after multiplication by a 
suitable power of $s$ shows that this is misleading (Fig.2b).  Even $P_{\rm int}(s)$ 
is not convex for $\theta > 1000$, and it develops an increasingly sharp shoulder 
near $s=s_{\rm max}$ as $\theta \to\infty$. We can try to obtain an alternative 
estimate of $\tau$ by fitting straight lines such that they touch both maxima in 
Fig.2b. Results of this are shown in Fig.3. In contrast to the previous estimate of 
$\tau$ they would indicate that the effective $\tau$ decreases with $\theta$ and is 
clearly less than 2.19 for $\theta\to\infty$ (also the value 2.159 of \cite{honecker}
seems hardly compatible with the extrapolation of Fig.3 to $\theta\to\infty$). But 
a convergence to $2.11$ seems quite possible, as we would expect if the bumps in 
fig.1b have a width independent of $\theta$ (which is not excluded by the data).

This obviously means that these scaling 
violations are not corrections which disappear in the limit $\theta\to\infty$, as 
was claimed in \cite{vespignani}. It rather indicates that the conventional scaling 
picture, and in particular the usual ansatz 
\be 
   P(s) = s^{1-\tau}\phi(s/s_{\rm max}),
\ee
is basically wrong, as was already conjectured in \cite{grass,schenk}.

\begin{figure}
\psfig{file=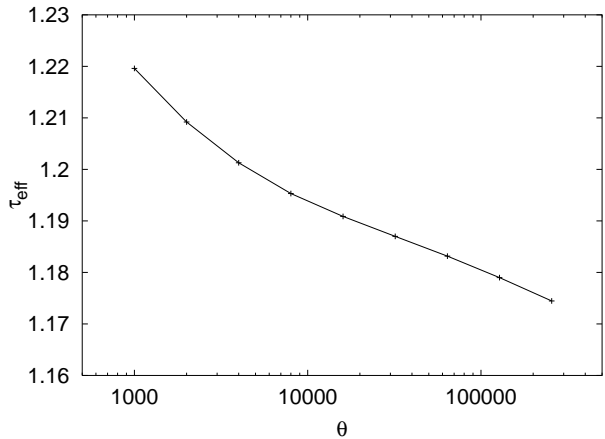,width=6.cm,angle=270}
\caption{Log-log plot of effective $\tau$ values, obtained by drawing lines 
   tangent to both maxima in Fig.2b. Notice that this is only defined when 
   $P_{\rm int}(s)$ is not convex, i.e. only for $\theta \ge 1000$.
   Error bars are roughly of the size of the symbols.}
\label{tau.ps}
\end{figure}

Previous analyses indicated that $s_{\rm max}$ actually increased faster than 
$\theta$, roughly as 
\be
   s_{\rm max} \sim \theta^\lambda               \label{lambda}
\ee
with $\lambda\approx 1.08$. We verified this qualitatively, but verified also the 
finding of \cite{grass} that $s_{\rm max}$ and thus also $\lambda$ are not well 
defined since the sharpness of the cut-off increases with $\theta$. This is already 
obvious from Fig.2b, but it persists for larger values of $s$ not shown in this 
plot.  It can also be seen by making copies of Fig.2a on transparencies and 
overlaying them.

Better scaling than for $P(s)$ was seen in \cite{grass} for $R(\theta)$. Our present 
data are fully compatible with those of \cite{grass,clar,honecker,footnote1}, but 
involve much higher statistics and cover a wider range of $\theta$. Thus it might not 
be too surprising that we now do not see perfect scaling any more. But the observed 
deviations from a pure power law (see Fig.4) are much larger than expected from 
subasymptotic corrections \cite{footnote3}. They clearly show that the previously 
seen power law was spurious and does not describe the asymptotic behaviour. A power 
law fit through the last two points would give $R(\theta)\sim \theta^{0.563}$, but 
this is obviously not yet the asymptotic behaviour. Again, as for $P(s)$ and 
$P_{\rm int}(s)$, we cannot yet say what the correct asymptotic behaviour will be.
In any case, the claim of \cite{honecker} that there are two different diverging
correlation lengths becomes obsolete.

\begin{figure}
\psfig{file=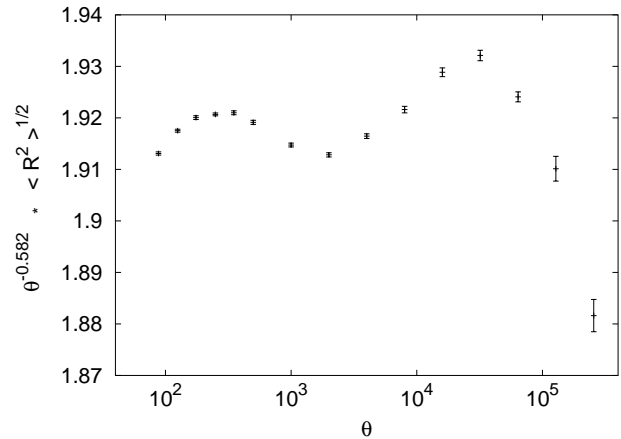,width=6.cm,angle=270}
\caption{Log-linear plot of $\theta^{0.582}\, R(\theta)$ versus $\theta$,
   where the power of $\theta$ was chosen such that the data fall roughly on a 
   horizontal line. Without the factor $\theta^{0.582}$, the data would be 
   hardly distinguishable from a straight line on a log-log plot.}
\label{nu.ps}
\end{figure}

\begin{figure}
\psfig{file=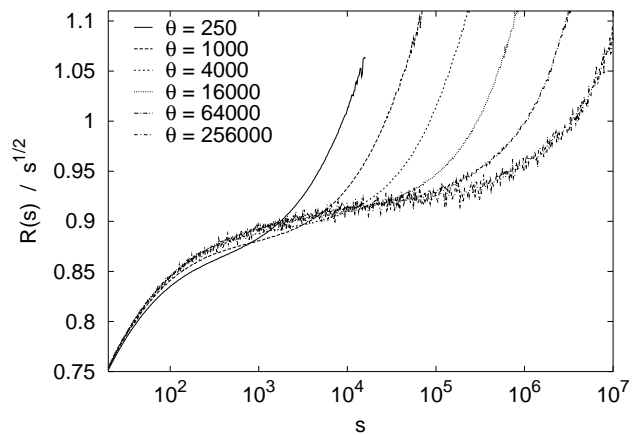,width=6.cm,angle=270}
\caption{Log-linear plot of $R(s,\theta) / \sqrt{s}$ versus $s$, for selected
   values of $\theta$.}
\label{rs.ps}
\end{figure}

\begin{figure}
\psfig{file=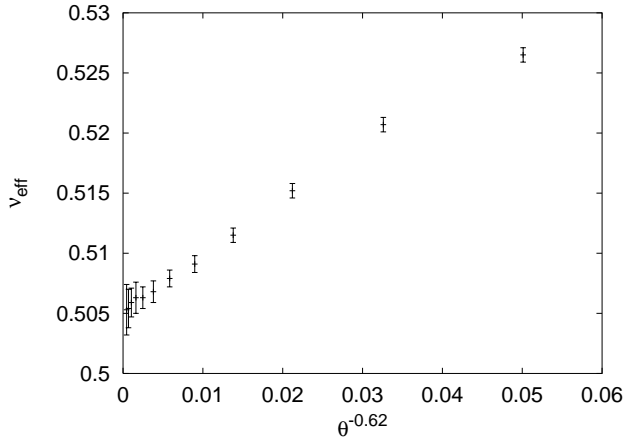,width=6.cm,angle=270}
\caption{Plot of $\nu_{\rm eff}$ versus $1/\theta$, where $\nu_{\rm eff}$ is 
   defined by $R(s,\theta)\sim s^{\nu_{\rm eff}}$ at the inflection points in Fig.5.}
\label{nuu.ps}
\end{figure}

An indication of the origin of all these puzzles comes from looking at $R(s,\theta)$.
This is defined analogously to $R(\theta)$, except that the averaging is now done 
over all fires of fixed size $s$. We expect $R(s,\theta) \sim s^{1/D}$ if fires are 
fractal with dimension $D$. Previous analyses had given $D\approx 2$ (or slightly
less). Thus we do not plot $R(s,\theta)$ versus $s$, but $R(s,\theta)/s^{1/2}$. 
From Fig.5 we see that there are 3 distinct regions which by and large coincide
with the three regions (left bump, flat middle part, and right peak) in Fig.1b.
\begin{itemize}
\item Region I: These are very small clusters, of size $s<300$. For them $R(s,\theta)$
increases faster than $\sqrt{s}$, which might suggest $D<2$. But we rather interpret
this as a finite-cluster artifact. Indeed, even compact clusters (with $D=2$) with 
a crumpled boundary will show an effective fractal dimension $<2$. These small 
clusters arise from lightnings which hit either very small regions with supercritical
tree density which burn off completely, or regions of subcritical density on which 
the fires form subcritical percolation clusters. Both mechanisms give compact clusters 
with fuzzy boundary. In any case, region I becomes less and less important as $\theta$
increases. \\
\item Region II: This is roughly equal to the scaling region in fig.1b. Here $R(s)$ is 
nearly proportional to $\sqrt{s}$, i.e. $D$ is very close 
to 2 -- but not quite. Also, we see a clear decrease of the minimal slope in Fig.5
with $\theta$, but it seems not to be sufficient to give $D=2$ in the limit $\theta\to
\infty$ (see Fig.6). According to Fig.6 the critical exponent $\nu$ defined by 
$R(s,\theta) \sim s^\nu$ in region II seems to converge to 0.505 for $\theta\to\infty$,
corresponding to $D = 1.98$. This is half way between the best previous estimates
$D=1.96$ \cite{clar} and $D=2.0$ \cite{grass}. It indicates that clusters in Region
II are fractal, but more compact than critical percolation clusters ($D= 1.89$).\\
\item Region III: In the region of very large fires, corresponding to the increasing 
bumps in Fig.1b, the apparent fractal dimension decreases again. Unfortunately, due
to the strong curvatures of the curves in Fig.5 in this region, we cannot quote 
any definite dimension value. But by plotting $R(s,\theta)/s^\nu$ with suitable 
values of $\nu$ on log-log scales, we see that (i) the maximal values of $\nu$ 
increase slowly with $\theta$ (except for the very largest $\theta$ where statistics 
is poor); and (ii) for our largest $\theta$ and $s$, we have 
$\nu \approx 0.64$. Thus the largest fires are definitely more fractal than critical
percolation clusters!
\end{itemize}

In terms of the scenario with roughly homogeneous patches with constant tree 
density \cite{grass,clar,schenk}, fires in region II correspond to single patches 
of typical size which are hit by lightning just when their density has reached 
about the critical percolation threshold. If lightning would always strike 
exactly at criticality, and if all trees in the patch would have burnt during the 
last fire so that all trees now are placed randomly, this would give $D=1.89$. Fires
in region II are more compact mainly because it will take some time until a 
lightning strikes by chance the region, after it has reached the critical density.

The larger a patch is, the bigger will be its perimeter and thus also the chance
that a fire ``spills over" and burns also a neighbouring patch, and from this 
also a next patch, etc. Since this will happen only along parts of the perimeter,
the resulting fires will be rather fuzzy, with effective dimension $<2$. We propose
that such fires dominate in region III. Although they are rather few in number, they
are very important since they burn large parts of the entire lattice, and they 
lead to rearrangements of the global pattern of patches. Notice that fires in region 
III burn only small parts of the entire region they cover, leaving behind more 
unburnt trees than fires of type II. Since they dominate more and more as 
$\theta$ increases, one might suspect that this leads to an increase of $\rho$
with $\theta$.

\begin{figure}
\psfig{file=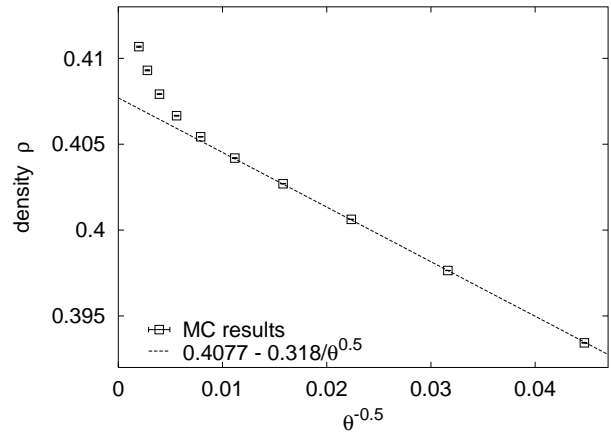,width=6.cm,angle=270}
\caption{Plot of $\rho$ versus $1/\sqrt{\theta}$. These values were extrapolated 
   to $L\to\infty$ using Eq.(1), but for the largest $L$ the extrapolation shifted
   them by less than the error bars (which are indicated {\it inside} the squares).
   The dashed line is a fit of the small
   $\theta$ values ($\theta \le 8000$) to an ansatz $\rho = a-b/ \sqrt{\theta}$.}
\label{rho.ps}
\end{figure}

Indeed, a slight increase of $\rho$ with $\theta$ had been seen in all previous 
simulations. The best previous estimates were $\rho = 0.4075 - const / \sqrt{\theta}$ 
\cite{grass} (unfortunately, the constant multiplying $\theta^{-1/2}$ was 
estimated wrongly in \cite{grass} due to a simple mistake) and 
$\rho = (0.4084\pm 0.0005)-const/\theta^{0.47}$ \cite{honecker,vespignani}. 
The results of our present simulations are shown in Fig.7. For
$\theta < 10^4$ we see a perfect agreement with previous results, but for larger
$\theta$ there is dramatic disagreement: Our measured values are higher than 
predicted by extrapolation from small $\theta$, by up to 100 standard deviations.
It is not clear why this was missed in \cite{vespignani}. But we might mention that 
no data for $\theta>10000$ are shown in Fig.1 of \cite{vespignani}, although the 
authors claimed to have made high statistics measurements up to $\theta=32768$.

When plotted against $\log\theta$, our values of $\rho$ follow roughly a straight 
line for $\theta \ge 4000$. We should of course not take this as the asymptotic 
behaviour, since $\rho$ can never increase beyond $p_c = 0.5927\ldots$ which is 
the critical value for site percolation in 2 dimensions. But we can use it to 
obtain a very crude estimate of the order of magnitude when $\rho\approx p_c$ 
should be reached. It is $\theta \approx 10^{40}$.

Although any prediction based on such a large extrapolation should be taken with 
great care, we conjecture that indeed $\rho$ converges to $p_c$ for $\theta\to
\infty$. The main reason is that we see no other plausible scenario compatible with 
our present numerics. It is not clear whether fires in this limit correspond to 
critical percolation clusters, since we must expect that weak correlations in the 
tree densities survive in this limit, sufficiently so to spoil any agreement with 
uncorrelated percolation on large scales.

\begin{figure}
\psfig{file=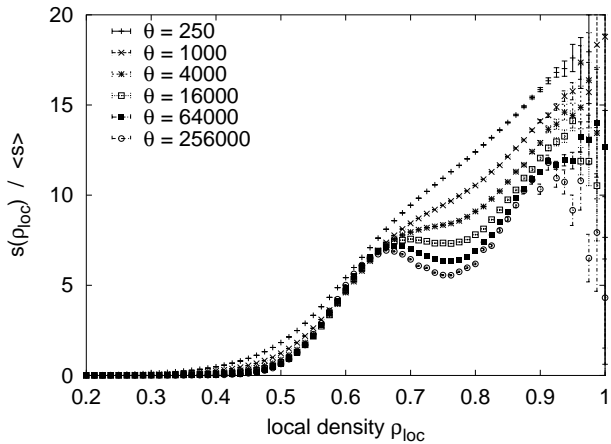,width=6.cm,angle=270}
\caption{Rescaled average fire sizes for fixed local density.}
\label{s_loc.ps}
\end{figure}

A last hint in favour of our claim that large fires asymptotically are 
dominantly associated to regions of critical tree densities (i.e., $\rho 
\approx p_c$), even though they might not form critical percolation clusters, 
is obtained by studying the mean sizes of fires which started in regions 
of given tree density. If a lightning hits a tree at site $i$, we define 
the local tree density $\rho_{\rm loc}$ at this site as the number of other 
trees in the surrounding square of size $9\times 9$ divided by 80. This size 
is of course rather arbitrary, but we obtained qualitatively similar results
for squares of sizes $7\times 7$ and $5\times 5$. In Fig.8 we plot the average
fire size $s(\rho_{\rm loc})$, divided by the overall mean $\langle s\rangle$, 
against $\rho_{\rm loc}$. For small $\theta$ we see a monotonic increase 
which is easily understood: Since even the largest patches of uniform density 
are not much larger than the square, large fires can only result from regions 
with high local density. This is no longer true for large $\theta$. There, patches 
with very high density are probably small, otherwise they would already have 
been burnt down earlier. At values of $\theta$ reached in this work the largest 
$s(\rho_{\rm loc})$ are still for large $\rho_{\rm loc}$, but a pronounced peak at
$\rho\approx p_c$ develops where $s(\rho_{\rm loc})$ has a local maximum. Since 
also the number of sites with $\rho_{\rm loc}\approx p_c$ is much larger than 
those with $\rho_{\rm loc}\gg p_c$ (see next paragraph), we see that it is fires 
in regions with critical density which play an increasing important role as 
$\theta\to\infty$.  In any case, the very strong dependence of $s(\rho_{\rm loc})$ 
on $\theta$ shows that we are still far from the asymptotic region where we 
expect this dependence to disappear.

\begin{figure}
\psfig{file=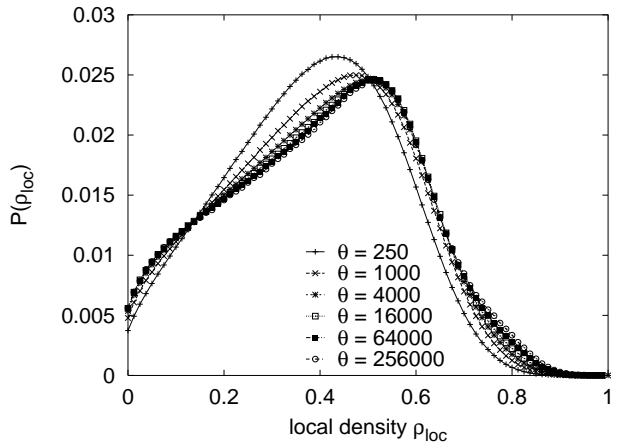,width=6.cm,angle=270}
\caption{Distribution of local densities.}
\label{P_loc.ps}
\end{figure}

Finally, we show in Fig.9 the distribution of local densities itself. It depends
rather weakly on $\theta$ (less than $s(\rho_{\rm loc})$, at least), but the 
precise way it does depend on $\theta$ is rather surprising and not yet fully
understood. First of all, it develops an increasingly sharp maximum which slowly 
shifts from $\rho_{\rm loc}\approx 0.4$ to $\rho_{\rm loc}\approx 0.6$ as 
$\theta$ increases. This is to be expected after our previous observations. What 
is unexpected and hard to explain is a shoulder at  $\rho_{\rm loc}\approx 0.8$
which develops for the largest $\theta$ values. It is not very large but 
statistically highly significant (it was seen in all runs with $\theta \ge 32000$).
Presumably related to it is a shoulder at small $\rho_{\rm loc}$ ($\approx 0.1$) 
which also increases with $\theta$: If patches with density $> 0.7$ burn, they 
leave behind extremely strongly depleted patches. One possible reason why such 
high density patches can survive at all is that many large fires are fractal 
and leave behind small disconnected regions of fairly high but subcritical 
density. These regions then are too small to have a chance to be hit by 
lightning until their density has grown far beyond $p_c$.

\section{Discussion}

The simulations reported in this paper leave little doubt that all scenarios 
proposed so far for the Drossel-Schwabl forest fire model do not 
describe the asymptotic behaviour for $\theta\to\infty$, where the model
should show SOC according to the standard folklore. Indeed, we do not see 
much indications for {\it any} power laws in this model, as all proposed 
scaling laws seem to be just transient. There are a number of observables which 
do show straight lines on log-log plots (such as Fig.1a in the central region
of $s$ or the envelope to Fig.1a), but it seems more likely that also these are 
spurious.

This situation is of course not altogether new. There are a number of models 
which were supposed to show anomalous scaling, but closer scrutiny proved 
otherwise. A good example is the Bak-Chen-Tang forest fire model \cite{bct}
which at least in $d=2$ is not critical \cite{gk}. Other examples include
the Newman-Sneppen model \cite{newman} where one can prove semi-numerically 
that the observed power laws are transient \cite{grass-unpub} and maybe even the 
``classical" abelian sandpile model in $d=2$. While power laws for {\it waves} 
were proven rigorously in that model, it might well be that the observed 
deviations from finite size scaling \cite{demenech,tebaldi} do not herald 
multifractality but just simply no scaling at all. One indication for the 
latter is the fact that some scaling laws show violations which do not seem 
do vanish for increasing system sizes \cite{glkp}. Also, some other quantities 
in the sandpile model which involve superpositions of many waves depend {\it 
qualitatively} on the geometry of the lattice (square {\it vs.} strip)
\cite{grass-unpub}. For a system with true scaling one would not expect this.

The situation becomes even worse when going to real life phenomena. It does not 
seem unlikely that many of the observed scaling laws are just artifacts or 
transients. Problems of spurious scaling in models which can be simulated with 
very high precision such as the present one should be warnings that not every 
power law supposedly seen in nature is a real power law.

Acknowledgement: I am indebted to Walter Nadler for critically reading the 
manuscript. 

Note added: After finishing this paper, I was informed that similar 
conclusions had been reached in a recent paper \cite{pruessner}, where 
the authors studied $P(s)$ with comparable (or even larger) statistics but 
on somewhat smaller lattices and for somewhat smaller $\theta$ than in the 
present paper. Unfortunately, neither the spatial sizes of fires nor 
precise values of $\rho(\theta)$ were measured in \cite{pruessner}. I am 
indebted to Gunnar Pruessner for sending me this paper, for very helpful 
correspondence and for pointing out some misprints in my manuscript.

\begin{table*}
\begin{center}
\begin{tabular}{|c|c|r|c|r@{ $\pm$ }l|r@{ $\pm$ }l|} \hline
 $\qquad\theta\qquad$ & $\;\;\log_2L\;\;$ & $N\qquad$ & $\quad N_{trans}\quad$ 
                  &\multicolumn{2}{c|}{density} &\multicolumn{2}{c|}{$R(\theta)$}\\ \hline
 256000   &  16 &$   9.3\times 10^6$ & $2.0\times 10^6$ & 0.410667 & 0.000036 & 2635.138&2.59   \\ \hline
 128000   &  16 &$   7.5\times 10^6$ & $3.0\times 10^6$ & 0.409321 & 0.000042 & 1789.675&1.96   \\ 
          &  15 &$  15.2\times 10^6$ & $1.6\times 10^6$ & 0.409231 & 0.000043 & 1787.024&1.37   \\ \hline
  64000   &  16 &$  46.4\times 10^6$ & $6.0\times 10^6$ & 0.407908 & 0.000013 & 1205.893&0.55   \\ 
          &  15 &$  42.5\times 10^6$ & $3.0\times 10^6$ & 0.407904 & 0.000037 & 1203.097&0.56   \\ 
          &  14 &$  37.7\times 10^6$ & $1.0\times 10^6$ & 0.407832 & 0.000038 & 1202.836&0.59   \\ \hline
  32000   &  15 &$  42.3\times 10^6$ & $6.0\times 10^6$ & 0.406623 & 0.000038 &  808.948&0.37   \\ 
          &  14 &$  78.3\times 10^6$ & $3.0\times 10^6$ & 0.406633 & 0.000037 &  807.253&0.27   \\ \hline
  16000   &  15 &$ 106.8\times 10^6$ & $8.0\times 10^6$ & 0.405428 & 0.000009 &  539.719&0.16   \\ 
          &  14 &$ 103.3\times 10^6$ & $5.0\times 10^6$ & 0.405401 & 0.000017 &  539.063&0.16   \\ 
          &  13 &$  73.2\times 10^6$ & $2.0\times 10^6$ & 0.405360 & 0.000029 &  538.464&0.20   \\ \hline
   8000   &  14 &$ 166.3\times 10^6$ & $6.0\times 10^6$ & 0.404188 & 0.000008 &  359.049&0.084  \\ 
          &  13 &$ 110.9\times 10^6$ & $3.0\times 10^6$ & 0.404135 & 0.000028 &  358.703&0.102  \\ \hline
   4000   &  14 &$ 207.8\times 10^6$ & $8.0\times 10^6$ & 0.402690 & 0.000007 &  239.213&0.050  \\ 
          &  13 &$ 394.2\times 10^6$ & $4.0\times 10^6$ & 0.402691 & 0.000010 &  239.207&0.041  \\ 
          &  12 &$ 332.7\times 10^6$ & $2.0\times 10^6$ & 0.402595 & 0.000011 &  238.992&0.044  \\ \hline
   2000   &  13 &$ 366.4\times 10^6$ & $6.0\times 10^6$ & 0.400614 & 0.000005 &  159.511&0.026  \\ 
          &  12 &$ 525.7\times 10^6$ & $2.0\times 10^6$ & 0.400577 & 0.000012 &  159.441&0.021  \\ \hline
   1000   &  13 &$ 319.6\times 10^6$ & $9.0\times 10^6$ & 0.397636 & 0.000005 &  106.647&0.018  \\ 
          &  12 &$ 458.4\times 10^6$ & $6.0\times 10^6$ & 0.397624 & 0.000006 &  106.681&0.015  \\ 
          &  11 &$ 655.8\times 10^6$ & $4.0\times 10^6$ & 0.397570 & 0.000010 &  106.603&0.013  \\ \hline
    500   &  12 &$ 490.8\times 10^6$ & $6.0\times 10^6$ & 0.393419 & 0.000006 &   71.414&0.0097 \\ \hline
          &  11 &$ 664.9\times 10^6$ & $3.0\times 10^6$ & 0.393399 & 0.000010 &   71.420&0.0083 \\ \hline
    350   &  12 &$ 663.4\times 10^6$ & $8.0\times 10^6$ & 0.390667 & 0.000004 &   58.093&0.0068 \\ \hline
    250   &  12 &$1000.0\times 10^6$ & $9.0\times 10^6$ & 0.387643 & 0.000003 &   47.757&0.0045 \\ 
          &  11 &$ 664.4\times 10^6$ & $6.0\times 10^6$ & 0.387626 & 0.000007 &   47.768&0.0056 \\ 
          &  10 &$1000.0\times 10^6$ & $2.0\times 10^6$ & 0.387574 & 0.000008 &   47.732&0.0045 \\ \hline
    175   &  11 &$ 733.6\times 10^6$ & $6.0\times 10^6$ & 0.383909 & 0.000006 &   38.788&0.0043 \\ \hline
    125   &  11 &$ 922.7\times 10^6$ & $8.0\times 10^6$ & 0.379837 & 0.000006 &   31.847&0.0032 \\ 
          &  10 &$1000.0\times 10^6$ & $3.0\times 10^6$ & 0.379804 & 0.000008 &   31.850&0.0030 \\ \hline
     88   &  11 &$1000.0\times 10^6$ &$12.0\times 10^6$ & 0.374935 & 0.000004 &   25.903&0.0025 \\ 
          &  10 &$1000.0\times 10^6$ & $8.0\times 10^6$ & 0.374910 & 0.000005 &   25.906&0.0025 \\ \hline
\end{tabular}
\caption{Statistics and main results: $N$ is the number of lightnings used for averaging, 
$N_{trans}$ that of lightnings discarded during the transients. The density is measured after
fires got extinct and before new trees are grown. Notice that the density has much larger
errors in some runs than in others of comparable statistics. This results from the fact that 
there are important long-ranged {\it negative} autocorrelations in the density time series, 
and I had not written out the information needed to take them into account in these runs.}
\end{center}
\end{table*}


\begin{thebibliography}{99}
\bibitem{footnote2} The burning times quoted in \cite{grass} were obtained from the 
   stack hights in a recursive (depth-first) implementation. They do not agree with 
   the more realistic burning times obtained with breadth first algorithms and should 
   thus be considered as obsolete.
\bibitem{footnote3} These deviations from scaling are qualitatively similar to those 
   seen on closer inspection in Fig.1 of \cite{honecker} for the correlation length 
   $\xi$, but not for $\xi_c$ which should be more closely related to $R(\theta)$.
\end{thebibliography}
\end{document}